\documentclass[twocolumn]{aastex63}
\usepackage{hyperref}

\begin{document}

\title{An Unambiguous Separation of Gamma-Ray Bursts into Two Classes from Prompt Emission Alone}

\correspondingauthor{Christian K. Jespersen}
\email{christian.jespersen@nbi.ku.dk}

\author[0000-0002-8896-6496]{Christian K. Jespersen}
\affiliation{Cosmic Dawn Center (DAWN)}
\affiliation{Niels Bohr Institute, University of Copenhagen, Lyngbyvej 2, DK-2100 Copenhagen \O}

\author{Johann B. Severin}
\affiliation{Cosmic Dawn Center (DAWN)}
\affiliation{Niels Bohr Institute, University of Copenhagen, Lyngbyvej 2, DK-2100 Copenhagen \O}

\author[0000-0003-3780-6801]{Charles L. Steinhardt}
\affiliation{Cosmic Dawn Center (DAWN)}
\affiliation{Niels Bohr Institute, University of Copenhagen, Lyngbyvej 2, DK-2100 Copenhagen \O}

\author{Jonas Vinther}
\affiliation{Cosmic Dawn Center (DAWN)}
\affiliation{Niels Bohr Institute, University of Copenhagen, Lyngbyvej 2, DK-2100 Copenhagen \O}

\author{Johan P. U. Fynbo}
\affiliation{Cosmic Dawn Center (DAWN)}
\affiliation{Niels Bohr Institute, University of Copenhagen, Lyngbyvej 2, DK-2100 Copenhagen \O}

\author{Jonatan Selsing}
\affiliation{Cosmic Dawn Center (DAWN)}
\affiliation{Niels Bohr Institute, University of Copenhagen, Lyngbyvej 2, DK-2100 Copenhagen \O}

\author{Darach Watson}
\affiliation{Cosmic Dawn Center (DAWN)}
\affiliation{Niels Bohr Institute, University of Copenhagen, Lyngbyvej 2, DK-2100 Copenhagen \O}

\begin{abstract}
The duration of a gamma-ray burst (GRB) is a key indicator of its physics origin, with long bursts perhaps associated with the collapse of massive stars and short bursts with mergers of neutron stars. However, there is substantial overlap in the properties of both short and long GRBs and neither duration nor any other parameter so far considered completely separates the two groups. Here we unambiguously classify every GRB using a machine-learning, dimensionality-reduction algorithm, t-distributed stochastic neighborhood embedding (t-SNE), providing a catalog separating all \emph{Swift} GRBs into two groups.  Although the classification takes place only using prompt emission light curves, every burst with an associated supernova is found in the longer group and bursts with kilonovae in the short, suggesting along with the duration distributions that these two groups are truly long and short GRBs. Two bursts with a clear absence of a supernova belong to the longer class, indicating that these might have been direct-collapse black holes, a proposed phenomenon that may occur in the deaths of more massive stars.

\end{abstract}

    
\section{Introduction}
The duration of a $\gamma$-ray burst (GRB) is a key indicator of its origin, with long-duration bursts typically associated with the core-collapse of a stripped massive star \citep{hjorth2003,Stanek2003} and the short-duration bursts with mergers of neutron stars \citep{1GRBKN,berger2013,Ghirlanda2018}. The dividing line is usually set at $T_{90} \approx 2$\,s \citep{Kouveliotou1993,Tavani1998,Paciesas1999}. However the long and short distributions are known to overlap substantially and neither duration nor any other parameter so far considered such as spectral hardness \citep{Kouveliotou1993} or lag \citep{Norris1986,Norris2006} gives a clean separation of the progenitor type based on the prompt properties. 

Considerable effort has been expended in trying to find a clean separation of burst types based on many other burst properties \citep{06environmentfruchter,NAKARsgrb,Bromberg11,zhang2012revisiting, Bromberg_2013classification}.
However, a complete separation has not yet been possible.

Ideally, classifying bursts should be done from the entirety of the light curve. However, the full light curve is a high-dimensional dataset, which makes comparing bursts hard because it is difficult to determine which information is most important.  As a result, proposed classifications have previously relied on a small number of easily-described summary statistics such as duration, spectral hardness and lag.  This has led to significant progress, but not to a clean separation between short and long GRBs.  

It has been hoped that with the right choice of summary statistics, there would be a clear separation into short and long GRBs.  This work uses the dimensionality reduction algorithm t-distributed Stochastic neighbour Embedding (t-SNE; \citealt{maaten2008visualizing,vandermaaten2015}) to instead reduce full GRB light curves to points in a two-dimensional space.  The location each light curve is mapped to in these two coordinates does not represent properties of that light curve, but rather can only be calculated from information about all of the light curves in the entire dataset.  Applying this technique to light curve data from the \emph{Swift} satellite, we find a clear separation into two groups, with possible additional subgroups.

In \S~\ref{sec:method}, we describe t-SNE and the process of assembling a map from the \emph{Swift} dataset.  The resulting map and implied classification into short and long GRBs is shown in \S~\ref{sec:classification}, along with a comparison between each group with expected properties. In  \S~\ref{sec: Speciel_Interest} we discuss how this classification works for GRBs of special interest. These results are discussed further in \S~\ref{sec:discussion}, including an application of this categorization to GRBs whose classification has been the subject of past debate.  

\section{Methods: Applying t-SNE to GRB light curves}
\label{sec:method}

The classification proposed here takes the entirety of the normalized \emph{Swift} light curves from prompt emission and in an unsupervised way determines which GRBs should be considered similar based on the prompt data alone.  This is done using t-SNE, a dimensionality reduction algorithm which can take complex, high-dimensional data and produce a faithful representation of that data in a low-dimensional space.

\subsection{t-Distributed Stochastic Neighbour Embedding}

t-SNE takes a set of high dimensional vectors, $\{\vec{x}_i\}$, and calculates the probability that each $\vec{x}_j$ should be considered a neighbour of $\vec{x}_i$ from the set of Euclidean distances $\{|\vec{x}_j-\vec{x}_i|\}$ and the \textit{perplexity}, a hyperparameter which determines the sizes of the neighbourhoods based on the density of the data in the respective regions and can be approximately interpreted as the typical number of neighbours which should be considered similar when computing distances. \textit{Perplexity} is formally defined \citep{maaten2008visualizing} as
\begin{equation}
    \textrm{Perp}(P_i)=2^{H(P_i)},
\end{equation}
where $P_i$ represents the conditional probability distribution over all other $\{\vec{x}_j\}$ given $\{\vec{x}_i\}$, and $H(P_i)$ is the Shannon entropy of $P_i$ in bits.
A higher \textit{perplexity} increases the values for points further away and promotes global structure, whereas a lower \textit{perplexity} is used to look for local structures and groupings of only a few vectors. The introduction of \textit{perplexity} is necessary since dimensionality reduction cannot simultaneously preserve both small and large scale structure in lower dimensions, and perplexity controls their relative importance. It should be noted that generating probabilities from Euclidean distances requires that every $\vec{x}_i$ have identical dimensionality, with none missing. 

These probabilities are then used to map the set of $\{\vec{x}_i\} \rightarrow \{\vec{y}_i\}$ in the lower-dimensional space, such that the probability of $\vec{y}_i$ and $\vec{y}_j$ being neighbours in the new mapping is as similar as possible to the probability that $\vec{x}_i$ and $\vec{x}_j$ were considered neighbours. This process depends upon random initialization and running t-SNE on an identical dataset can produce a variety of maps with similar toplogy (cf. \citealt{Steinhardt_tsne_galaxy2020}, Figure 1).  Thus, the axes of this low-dimensional space have no proper labels or meaning; unlike principal component analysis, an object further to the right on a t-SNE map is not in some sense more `x-like', but merely more similar to nearby objects and less similar to more distant ones.
Essentially, the goal of t-SNE is to produce a simplified map which can be easily visualized while preserving the structure of the original dataset, not to find a vector basis.

Although t-SNE is considered an unsupervised algorithm, in practice the final map depends upon not the data and hyperparameters alone, but also a series of human selections in preparing that data.  Producing a dataset of vectors with identical components requires a combination of removing individual objects for which some properties are poorly measured or unmeasured and instead removing some properties from the calculation altogether (or choosing an alternative metric for calculating neighbour probabilities).  A choice of how to format and scale the raw data can change the relative weighting of various components in calculating neighbour probabilities.  These human influences can often be minimized by running t-SNE on highly standardized datasets, such as the catalog produced by one survey or one observatory.

\subsection{Data Preparation for the \emph{Swift} Catalog}
\label{subsec:dataprep}

Here, t-SNE is run on the full \emph{Swift} GRB catalog \footnote{available at https://swift.gsfc.nasa.gov/results/batgrbcat/} \citep{Lien2016} in an attempt to classify all detected bursts.  \emph{Swift} measures the light curve of each burst in four bands, 15--25\,keV, 25--50\,keV, 50--100\,keV, and 100--350\,keV. The \textit{Swift} data are released as four binned, background-subtracted light curves with temporal resolution 64\,ms, and the start of the burst may vary slightly from the trigger time.  Because bursts vary greatly in duration, there is a very wide range in the number of time bins with statistically significant flux, where the reported flux for \emph{Swift} is measured as photon count/cm$^2$/s.

Therefore, it is essential to standardise the dataset prior to running t-SNE, ideally in a manner which preserves meaningful differences but erases differences which should not have physical origin.  For example, a translation of all light curves by the same constant time offset should be ignored. A similar problem is tackled when analysing the benchmark  \textbf{M}odified \textbf{N}ational \textbf{I}nstitute of \textbf{S}tandards and \textbf{T}echnology (MNIST) handwritten digit dataset 
with machine learning, since the data needs a large amount of preprocessing (e.g. normalization, deskewing, noise removal) in order to be effectively analysed \citep{mnistlecun1998gradient, deng2012mnist}. The goal is therefore to prepare data for t-SNE in a way that retains all of the useful information but removes irrelevant information that the analysis might otherwise use for classification. The end result of this process must be a set of vectors of identical dimension, for which it is still proper to expect that bursts with similar physical origin will be similar when compared component by component.

In our data, the main possible distractions would be (1) the total time-integrated flux, which would only return the known result that long GRBs typically but not always have higher fluence; (2) different lengths of light curve measurements, which can depend upon noise in the tail of the \emph{Swift} light curves and is therefore extremely important; and (3) the possibility of a trigger time offset common to all four \emph{Swift} bands, although any relative time delay between different bands is still meaningful.  

To remove these, light curves are normalized in every band by the total time-integrated flux across all bands of that specific burst (removing (1)), zero-padded in order to ensure a common axis and length (partly removing (2)\footnote{The remainder of (2) is completely accounted for by the fact that an erroneous bias in duration would only, at worst, include noise-dominated measurements which do not affect the FT significantly, and do not add any significant fluence.}; cf. \citealt{FTopticszero,FTIEEEzero}),  then concatenated before taking the discrete-time Fourier transform (DTFT) of the light curve measured in each individual band (removing (3)). Because t-SNE relies on Euclidean distances, omitting these steps would group bursts almost solely by effects which are known not to produce a meaningful classification, e.g. omitting the normalization would effectively group bursts nearly solely by the magnitude of the flux, which would render only a known result; that long GRBs typically, but not always, have higher fluence than short GRBs \citep[e.g.,][]{Ghirlanda2009}, but instead here bursts are grouped using the entirety of the remaining information. 

It should be noted that the DTFT is not necessarily an optimal preprocessing solution for separating GRBs. Other preprocessing techniques which preserve the information content in the lightcurve would yield equally valid representations of GRB classes, though possibly not as cleanly separated. As explained in the previous paragraph, the main motivation for the DTFT is that it suppresses some of the biases in the light curves, while building on previous successful descriptors (duration, hardness, spectral lag) indicating that the general shape of the light curve could be a defining characteristic. 

The DTFT also does not formally guarantee that similar light curves end up as close neighbours, but rather that light curves with similar DTFT end up close together, as t-SNE acts on Fourier components rather than flux measurements. Although in principle one could imagine constructing a function which would render a DTFT similar to a GRB light curve despite looking dissimilar, this is unlikely to occur in real data. Thus, having similar DTFTs is equivalent to having similar shapes, since the DTFT in essence is just a change in basis.

Previous attempts at classification have typically focused on summary data, such as duration (typically $T_{90}$, the duration containing 90\% of the statistically-significant flux, or $T_{100}$, containing 100\% of the flux), spectral lag, flux or fluence, and hardness.  These are properties which can be calculated from the light curves in each band, and therefore part of what is available to t-SNE, but represent only a small subset of the full dataset.  Here, the full \emph{Swift} light curves are encoded then used, and neighbour probabilities are ultimately produced from $\approx 30000$-dimensional vectors, encoding far more information than summary data.

\subsection{Summary of Classification Scheme}
\label{subsec:algorithm}

Using the background-subtracted light curves, \emph{Swift} bursts can be classified with the following procedure:
\begin{enumerate}
    \item {The 64 ms-binned light curve in each band is limited to the interval out to $T_{100}$.\footnote{The DTFT would be unaffected by the additional extra bins containing solely noise, and thus the limited information between $T_{90}$ and $T_{100}$ can be included.}
    Afterwards they are padded with zeros, placing light curves of different length on a standard basis. }
    \item {Individual light curves are normalized by the total fluence, obtained as the numerical integral of the flux across all bands, preserving spectral hardness}.
    \item {The resulting light curves are concatenated and the DTFT is applied, producing vectors suitable for input into t-SNE.}
    \item {t-SNE is applied to the vectors $x_i$, consisting of the Fourier amplitudes, with \textit{perplexity} set to 30 (a choice specific to the \emph{Swift} catalog), terminating once no improvement is made on the cost function.}
    \item {The resulting map is examined, in the hopes that bursts will be clearly separated into clusters which can be interpreted as arising from a similar physical origin.}
\end{enumerate}
The success of this procedure is evaluated in the following sections.

\section{Classification}
\label{sec:classification}

\begin{figure*}[!ht]
    \centering
    \includegraphics[width = \linewidth]{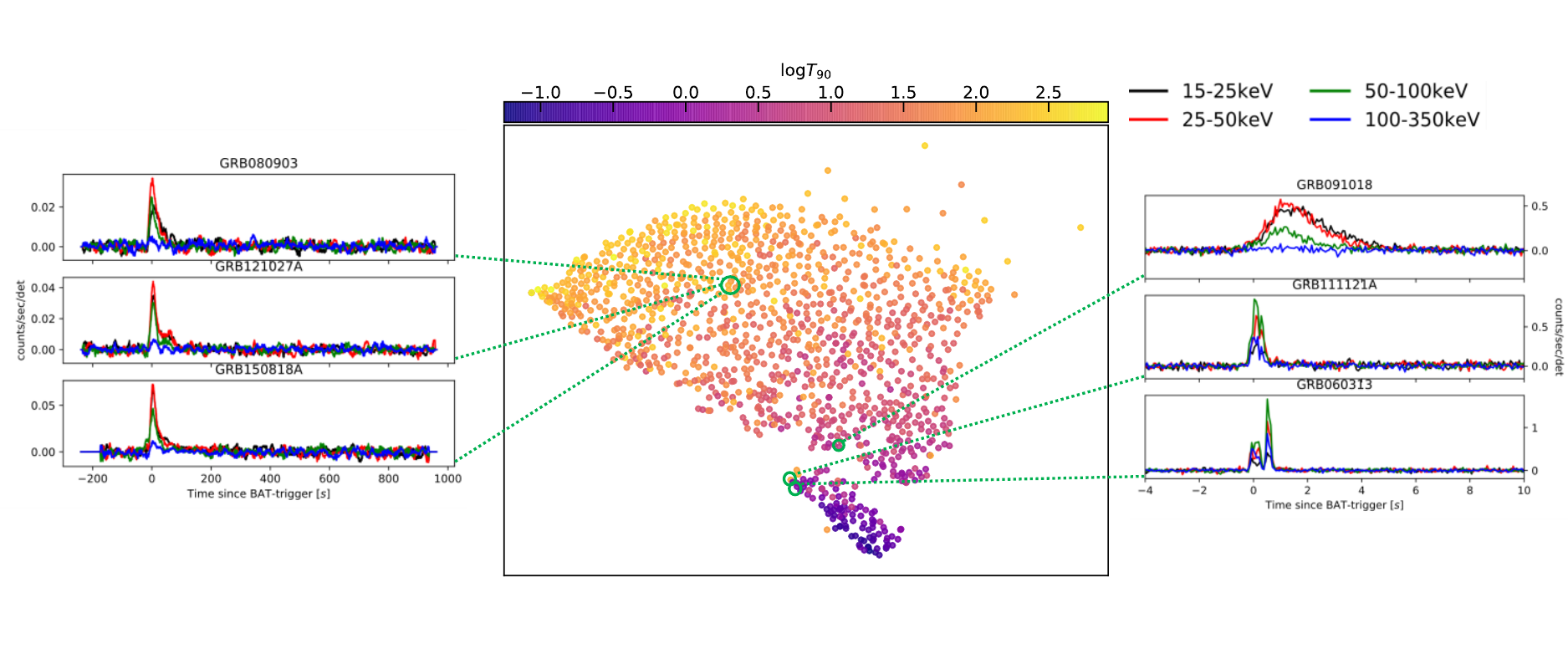}
    \caption{t-SNE mapping of \emph{Swift} light curves, colored based on duration $\log(T_{90})$. Several sample light curves in the four observed bands are shown, with similar light curves placed as near neighbours and dissimilar light curves placed further apart.  A clear separation into two groups is visible, with the smaller, bottom group (referred to as type-S) generally but not uniformly of shorter duration (see Fig. \ref{fig:resultfigs}). The axes resulting  from t-SNE have no clear physical interpretation or units; only the structure is meaningful.}
    \label{fig:main_figures}
\end{figure*}

The procedure described in \S~\ref{subsec:algorithm} indeed maps the GRBs in the \emph{Swift} catalogue to a two-dimensional space (Fig. \ref{fig:main_figures}). As a result, light curves with similar Fourier amplitudes end up as close neighbours, which translates to light curves with similar shape being grouped close together, as per the discussion in \S~ \ref{subsec:dataprep}. 

\begin{figure*}[!ht]
    \centering
    \gridline{\fig{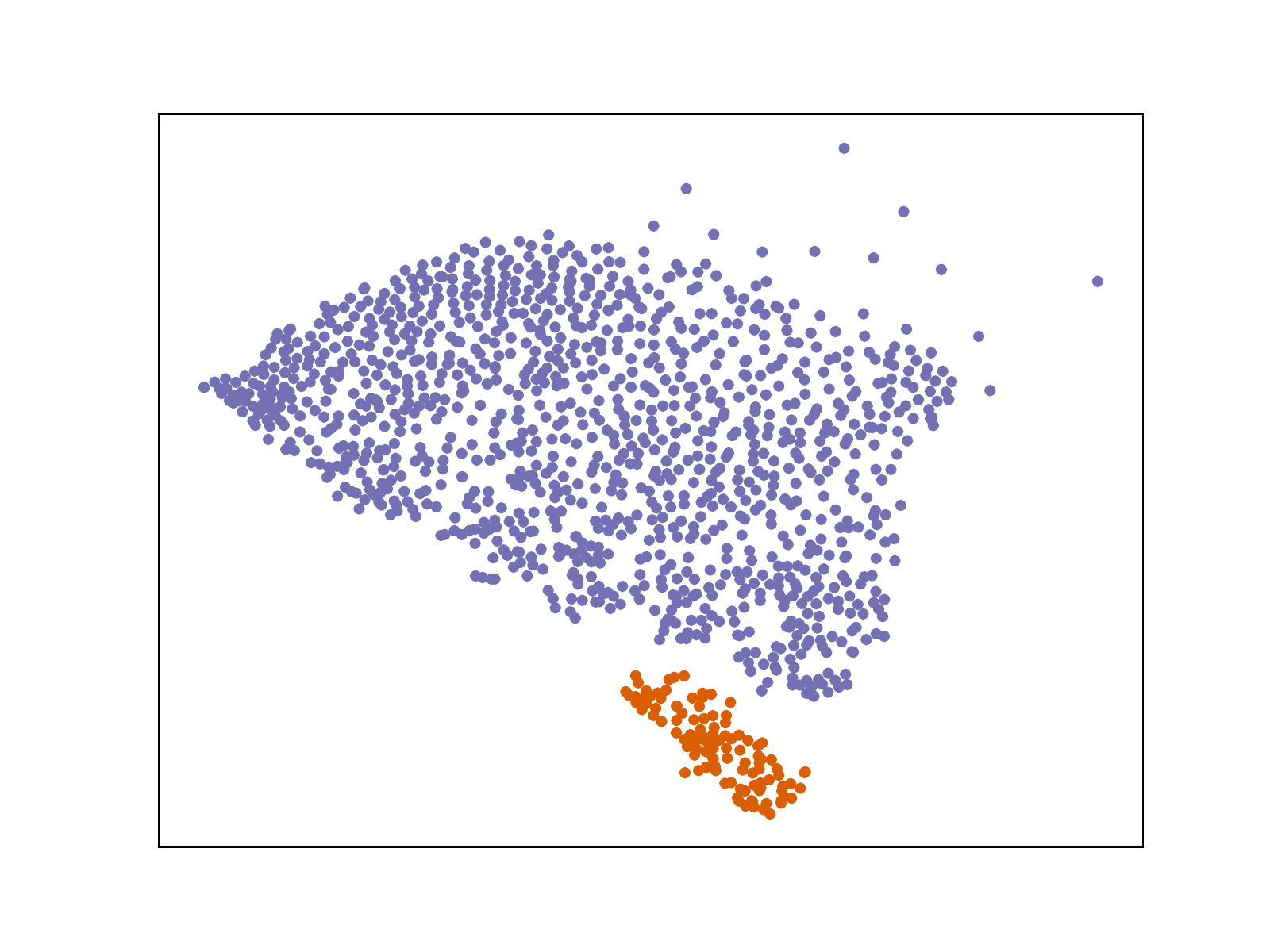}{0.5\linewidth }{(a)}
    \fig{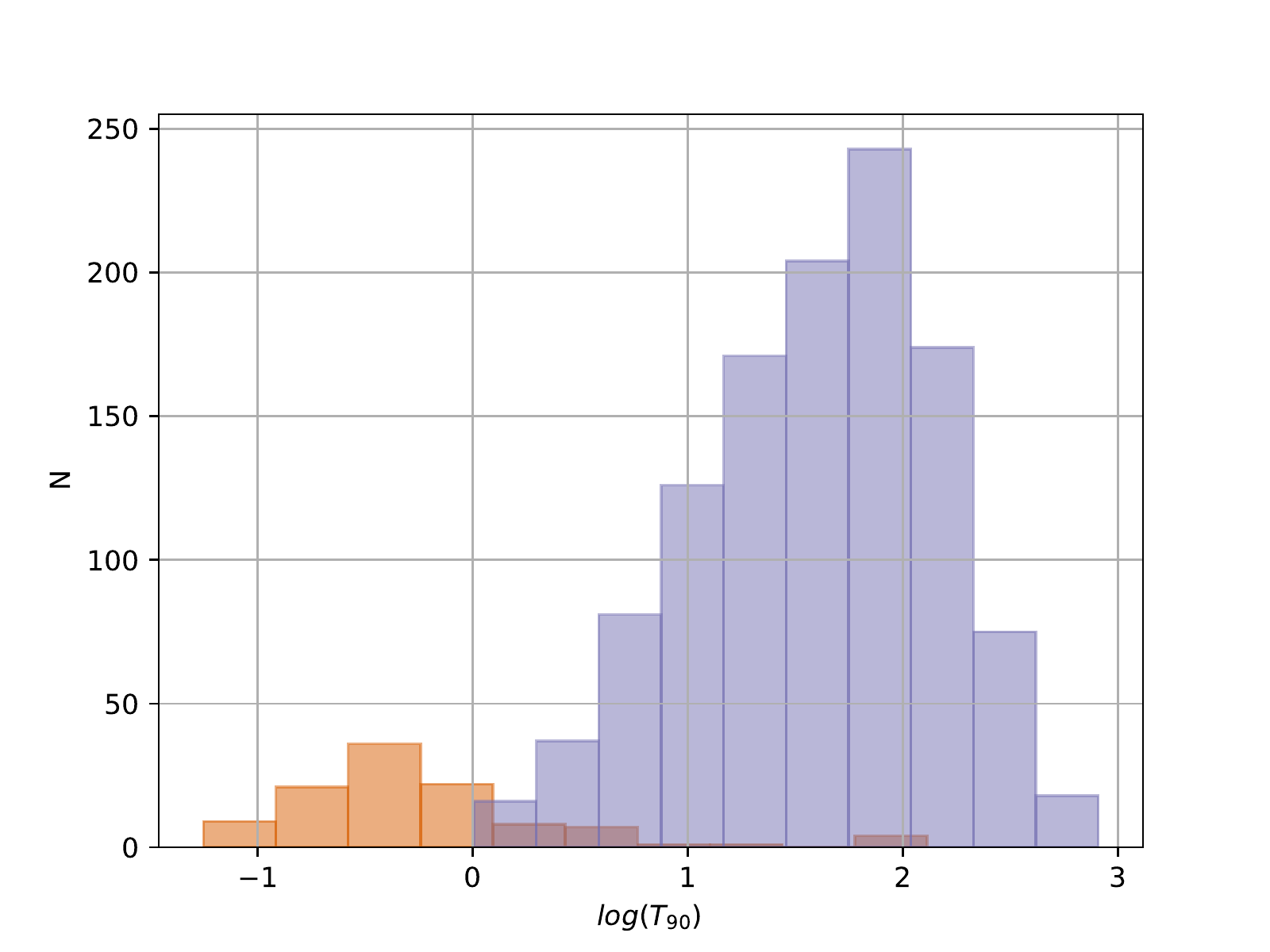}{0.5\linewidth }{(b)}}
    \caption{(a) The clear separation into two groups (purple and orange) strongly suggests a classification of GRBs into two distinct types. (b) The distributions of duration in log($T_{90}$) of type-L and type-S bursts are approximately normal and similar to those expected for a long- and short classification.}
    \label{fig:resultfigs}
\end{figure*}

Even more promisingly, there is discernible structure in the resulting map.  The exact mapping can change depending upon random seeds or the order in which data are presented to t-SNE, but structure such as grouping and topology remains (see Fig. 1 and related discussion in \citealt{Steinhardt_tsne_galaxy2020}). Bursts are divided into two groups, with a clear separation between the larger group at top and a smaller group at bottom as presented in Fig. \ref{fig:resultfigs}a. Further substructure exists as well, discussed in \S~\ref{subsec:comparison}.

A closer examination of the duration ($\textrm{T}_{90}$) of each burst indicates that the t-SNE map has grouped objects approximately (but not entirely) by duration (Fig. \ref{fig:main_figures}).  The larger group at top is generally of longer duration, while the smaller group at bottom is shorter.  Both groups have possibly (skewed) Gaussian distributions in log duration (Fig. \ref{fig:resultfigs}b), with some overlap in the $\textrm{T}_{90} \sim 1-10\textrm{ s}$ range.  

In short, the mapping cleanly divides \emph{Swift} GRBs into two groups, with the duration distributions of the two groups similar to what might be expected for a classification into short and long bursts. To avoid confusion with pre-existing classifications, the two t-SNE groups will be referred to as type-S and type-L.

A full list of our classification of \emph{Swift} GRBs is given in Table \ref{tab:classification}.
\begin{table}[] 

\begin{tabular}{l|ll}  

\textbf{GRB} & \textbf{$T_{90}$[s]} & \textbf{Type} \\ \hline
GRB\,190727B  & 39.2 & L          \\
GRB\,190719C & 185.8  & L          \\
GRB\,190718A  & 704.0   & L          \\
GRB\,190706B  & 43.6 & L          \\
GRB\,190701A  & 38.4 & L          \\
...         & ...    & ...           \\
\end{tabular}
\caption{Classification of \emph{Swift} GRBs as either type-S or type-L based on the separation in the DTFT-based t-SNE map.  A complete, machine-readable table is available online.}
\label{tab:classification}
\end{table}

\subsection{Comparison with Other Classifications}
\label{subsec:comparison}

Although the classification of many individual GRBs as short or long is uncertain, particularly for bursts of intermediate duration, many bursts can be unambiguously classified as short or long on the basis of other observations.  It is therefore necessary to confirm that the classification proposed here matches these previous results.

Perhaps the strongest association is that between supernovae and long GRBs, with every GRB with detection of an associated supernova unambiguously being long \citep{Hjorth2012,Cano2017}.  Eleven bursts in the \emph{Swift} catalog, GRB\,060218A, GRB\,071112C, GRB\,100316D, GRB\,111209A, GRB\,111228A, GRB\,120714B, GRB\,120729A, GRB\,130215A, GRB\,130831A, GRB\,161219B and GRB\,171010A have clearly detected associated supernova that are well-characterized by spectroscopy \citep{Bufano_2012, Cano2017, Cano_2014, Klose_2019, Kann_2019}. Due to a corrupted files for GRB\,060218A, not having BAT light curve data available for GRB071112C  \footnote{Since its emission overlapped with 071112B} and GRB171010A \footnote{GRB171010A is a Fermi burst that was followed up by \emph{Swift}}, these are not included.  The remaining eight are all classified as type-L by the t-SNE map (Fig. \ref{fig:interest}).

It has also been suggested that there may be a link between the hardness of the GRB spectrum and classification \citep{Kouveliotou1993, Ghirlanda2009}.  It is known that this does not produce a clean, unambiguous separation between short and long GRBs.  However, there is still likely a strong correlation between hardness and type, particularly since the hardness of a burst should be closely related to its physical origin.  
\begin{figure}[!ht]
    \centering
    \includegraphics[width=1\linewidth]{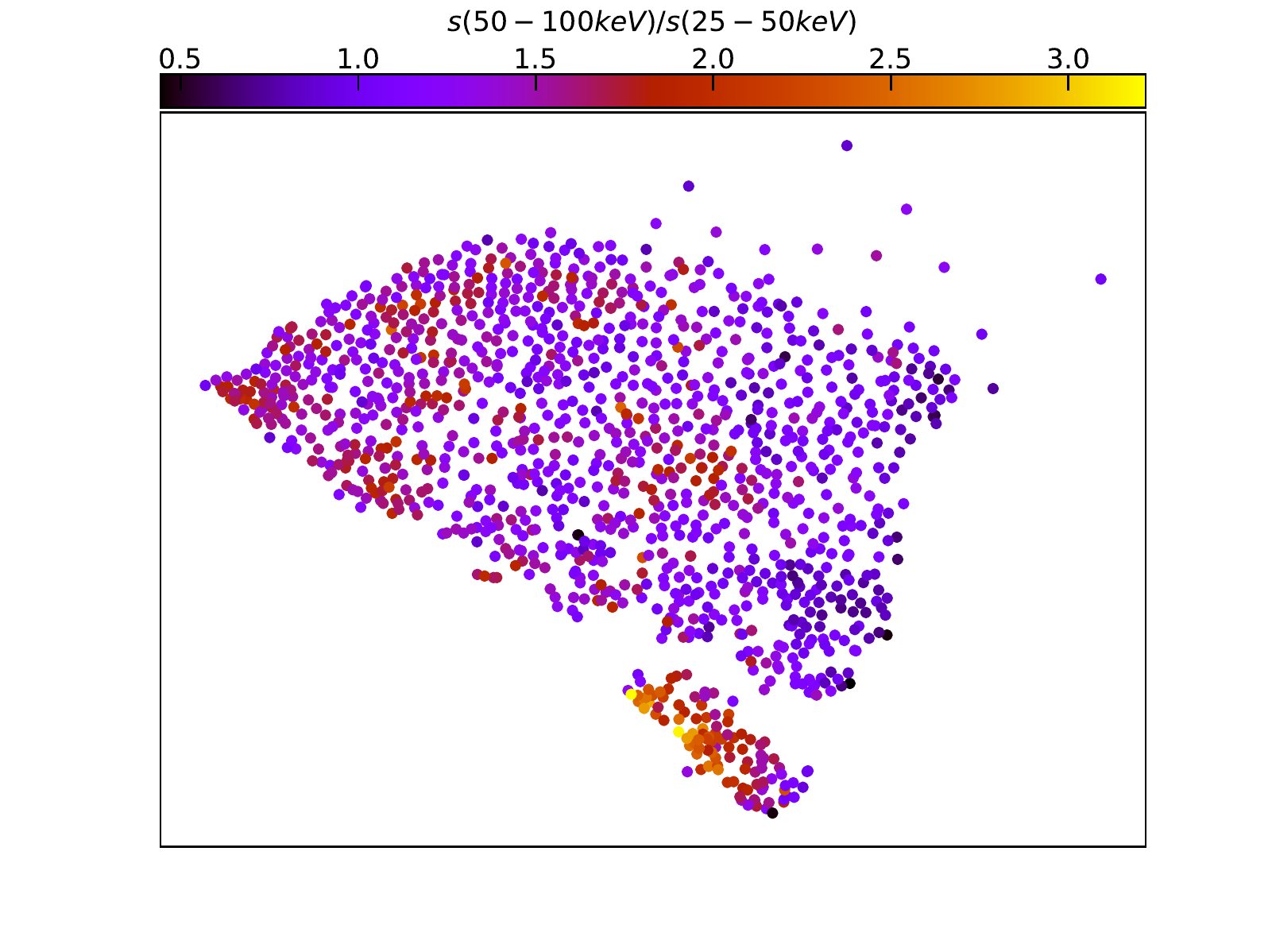}
    \caption{Colouring bursts on the t-SNE mapping by the flux
    ratio of the 50-100 keV and 25-50 keV bands, a proxy for hardness, indicates several clusters of hard bursts.  The hardest bursts are all classified as type-S, but some type-S bursts do not have hard spectra, and several tight clusters of type-L bursts also exhibit harder spectra.  A reasonable interpretation is that there are multiple classes both of type-S and L bursts with different physical origins, with some of these origins producing harder spectra.}
    \label{fig:hardnessandz}
\end{figure}

The t-SNE mapping indeed shows that bursts with similar light curves tend to have similar hardness (Fig. \ref{fig:hardnessandz}).  Bursts with harder spectra group together in tight clusters on the t-SNE map, and the hardest bursts lie in a tight cluster and are all categorized as type-S.  However, clusters of hard bursts exist both in the type-S and type-L group.  In general, most type-S bursts are harder than most type-L bursts, but hardness alone is insufficient to determine whether a burst is type-S or type-L.  Rather, the t-SNE map (Fig. \ref{fig:interest}, bottom) at lower \textit{perplexity} suggests that there might be substructure in both the type-S and type-L classes. The cause for this is currently undetermined, but could be associated with spectral hardness. 
Another measure of this feature could be the peak energy in $\nu f(\nu)$, $E_{peak}$.  However, \emph{Swift} frequently is unable to reliably measure $E_{peak}$ due to the relatively low energy bandpass of \emph{Swift}-BAT.  As a result, there is less structure in maps colored by $E_{peak}$ than in maps using spectral hardness ratio compared across the \emph{Swift}-BAT bandpass, although hardness also is an incomplete diagnostic.

\section{GRBs of Special Interest}
\label{sec: Speciel_Interest}
The clear separation on the t-SNE map into two groups, namely type-S and type-L, provides labels for all GRBs, including ones which were previously difficult to classify.

In addition to providing this broad classification, there is considerable additional information provided by the t-SNE map.  Because each burst is placed close to other, similar bursts, it is possible to determine both whether an object is typical (many neighbours) or atypical (fewer neighbours, and often on the edges of a cluster).  If some bursts have additional information available such as an observed afterglow or host galaxy, it may be natural to ascribe similar properties to neighbouring bursts for which those observations are unavailable.  

A full discussion of the various groups indicated is beyond the scope of this paper.  However, here we briefly describe the t-SNE classification of several GRBs which have previously been the subject of debate, as different analysis techniques have disagreed on their proper categorization.
\begin{figure}[!ht]
    \centering
    \includegraphics[width=0.78\linewidth]{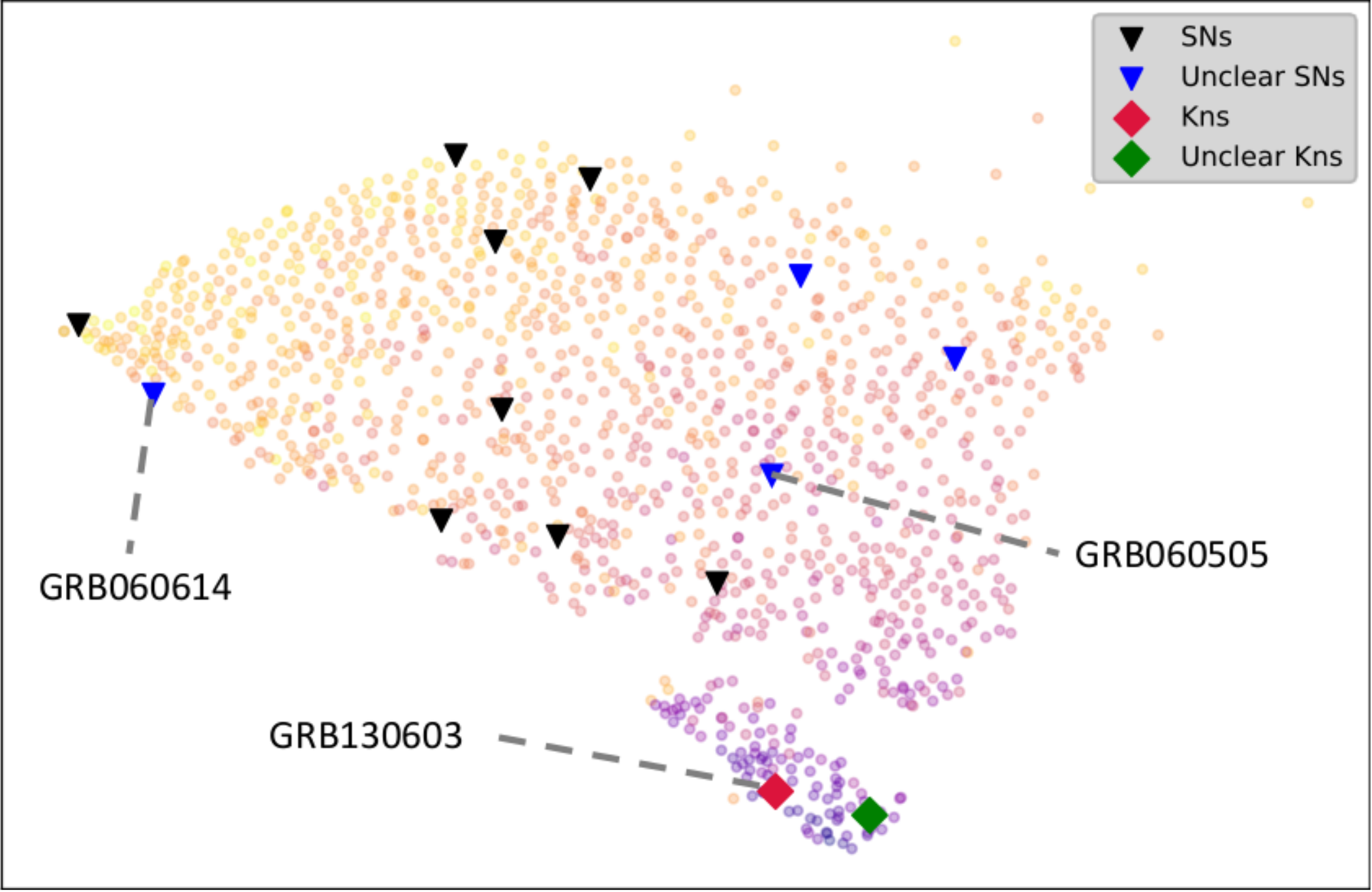}
    \includegraphics[width=1\linewidth]{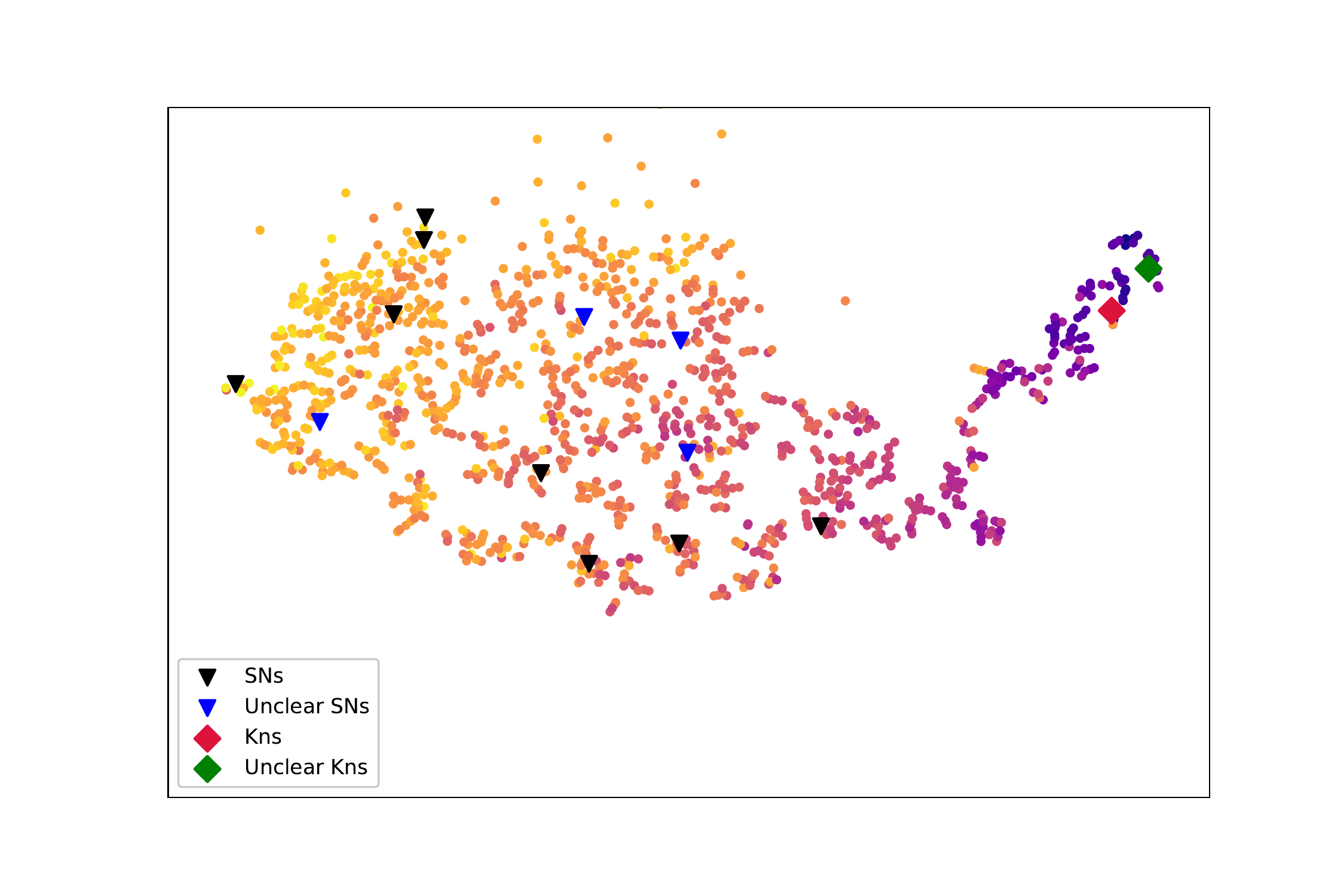}
    \caption{(top) The locations of several GRBs that have been the focus of recent studies are indicated on the t-SNE mapping used throughout.  Note that a complete list of the t-SNE classification into long and short GRBs is available in Table \ref{tab:classification}.  (bottom) Classification of GRBs using t-SNE run at a \textit{perplexity} of 5, lower than the map shown in the other figures.  The lower \textit{perplexity} is chosen to emphasize substructure.}
    \label{fig:interest}
\end{figure}

\subsection{Supernova-less long GRBs}

GRB\,060505 and GRB\,060614 are of special interest, since they are long duration GRBs with no observed optical counterpart \citep{DellaValle2006,NoSN2006Nature}. They have therefore been suggested to have their own progenitor mechanism, constituting their own class of Extended Emission Short Gamma Ray Bursts (EE sGRB; \citealt{Norris2006,Ofek2007}). 
The t-SNE map classifies GRB\,060614 as a type-L burst.  It lies close to GRB\,111209A, which has an associated luminous SN \citep{1SNGRB}, and in the same cluster as GRB\,111209A. This suggests that GRB\,060614 would not have a separate progenitor mechanism, but is instead a more standard long GRB. 

GRB\,060505 lies far away from the cluster of type-S GRBs and clusters with type-L GRBs. This indicates that GRB\,060505 was a long GRB. However, it doesn't group cleanly with the confirmed SNe, suggesting there may have been a different physical origin.

\subsection{Short GRB association with Kilonovae}

GRB\,130603B is a short GRB with a kilonova (KN) counterpart \citet{1GRBKN}. GRB\,130603B lies in the type-S GRB group close to GRB\,160821B, which has also been theorized to have had an associated KN \citep{2019MNRASKN, Lamb2019}. This suggests that GRB\,160821B may indeed be associated with a KN. \\
Another proposed KN candidate is GRB\,050509B, which is unfortunately removed from our catalogue due to having a duration of 1 time bin or less, making the DTFT meaningless. \\
Since only one burst with an optical KN counterpart is included in the catalogue, extrapolation is not very useful. As more KN are discovered, this method will become an increasingly important feature of the t-SNE map.  GRB170817A, the burst associated with GW170817 is not a \emph{Swift} burst and thus is not included in our analysis \citep{gw17, fermi170817A}.

\section{Discussion}
\label{sec:discussion}

The application of the dimensionality reduction algorithm t-SNE to GRB light curves observed by \emph{Swift} is used to group bursts based upon similarities with their neighbours.  The resulting map (Fig. \ref{fig:resultfigs}a) suggests that bursts should be classified into two broad groups.  An analysis of the duration distribution, optical counterparts, and previous proposed classifications strongly suggests that these two groups correspond to short and long GRBs.  If so, this technique would provide a complete, unambiguous separation of the GRB population into short and long bursts. A natural next step is to determine which properties best correlate with classification and use this to develop astrophysical models, although such a study is beyond the scope of this letter.

The t-SNE algorithm can be tuned to instead focus on substructure, and it is possible that further subgroups exist within these two populations (Fig. \ref{fig:interest}).  Substructure might be associated with distinctions between bursts belonging to these groups, perhaps indicating that a long GRB could be produced by multiple physical causes.  Because relatively few bursts have known counterparts or other additional information, it is difficult to connect these groups with physical origins.  However, the few bursts known to have a common origin, such as supernovae, are indeed mapped to nearby locations.  Similarly, the confirmed kilonova GRB\,130603B is a near neighbour of GRB\,160821B \citep{Lamb2019}, proposed as a kilonova candidate.

If these substructures indeed correspond with physical origin, then the t-SNE map can be used to select bursts for targeted followup.  For example, it would be straightforward to determine whether a new burst is a kilonova candidate based upon whether it is placed close to GRB\,130603B.  

With small modifications to avoid recalculating the entire map for a new object, t-SNE could also easily be used in an automatic classification pipeline along with a suitable clustering algorithm.  However, GRBs are sufficiently rare to warrant individual human attention, with good choice of parameters, a t-SNE map will have sufficient separation that human visual classification is both possible and likely preferable.

\subsection{Implications for anomalous GRBs}
This study demonstrates that GRB\,060614, long-debated as a possible extended emission short GRB, is indeed a long-duration type GRB. Its lack of a supernova down to 5 magnitudes below SN1998bw presents a real challenge to standard jet-driven supernova models and suggests a direct collapse black hole, as originally postulated by \citet{NoSN2006Nature}. The same is true for GRB\,060505, which was at least 6 magnitudes fainter than SN1998bw and is found unequivocally to belong to the long-duration category.  The most compelling interpretation would be that GRB\,060505 and GRB\,060614 were direct-collapse black holes \citep{Adams2017,Liu2019}, a possibility which has been theorized but not yet observed.

The methods developed in this work can be used to cleanly divide all \emph{Swift} light curves into two distinct classes.  Based upon the duration distribution and other properties it is tempting to label type-S and type-L GRBs as short and long.  However, this classification is entirely empirical, based upon the distribution of light curve properties rather than on astrophysics. One of the drawbacks of t-SNE is that it is not immediately clear which properties of the full light curves have been most influential in this classification or how those properties might relate to underlying GRB astrophysics.  Whether these indeed have distinct astrophysical origins, perhaps one arising from collapse and the other from collision, will therefore require additional followup studies of each group.  It is hoped that this process will be far more straightforward now that every GRB, even at duration $\sim 2$ seconds, can be unambiguously classified.

\section*{Acknowledgements}
The authors would like to thank Chryssa Kouvelitou, Jens Hjorth, Adam Jermyn, and Georgios Magdis for helpful discussions.  CLS is supported by ERC grant 648179 "ConTExt".  The Cosmic Dawn Center (DAWN) is funded by the Danish National Research Foundation under grant No. 140.

\bibliography{ref}

\begin{thebibliography}{}
\expandafter\ifx\csname natexlab\endcsname\relax\def\natexlab#1{#1}\fi
\providecommand{\url}[1]{\href{#1}{#1}}
\providecommand{\dodoi}[1]{doi:~\href{http://doi.org/#1}{\nolinkurl{#1}}}
\providecommand{\doeprint}[1]{\href{http://ascl.net/#1}{\nolinkurl{http://ascl.net/#1}}}
\providecommand{\doarXiv}[1]{\href{https://arxiv.org/abs/#1}{\nolinkurl{https://arxiv.org/abs/#1}}}

\bibitem[{Abbott {et~al.}(2017)Abbott, Abbott, Abbott, Acernese, Ackley, Adams,
  Adams, Addesso, Adhikari, Adya, \& et~al.}]{gw17}
Abbott, B.~P., Abbott, R., Abbott, T.~D., {et~al.} 2017, The Astrophysical
  Journal, 848, L12, \dodoi{10.3847/2041-8213/aa91c9}

\bibitem[{{Adams} {et~al.}(2017){Adams}, {Kochanek}, {Gerke}, {Stanek}, \&
  {Dai}}]{Adams2017}
{Adams}, S.~M., {Kochanek}, C.~S., {Gerke}, J.~R., {Stanek}, K.~Z., \& {Dai},
  X. 2017, \mnras, 468, 4968, \dodoi{10.1093/mnras/stx816}

\bibitem[{{Berger} {et~al.}(2013){Berger}, {Fong}, \& {Chornock}}]{berger2013}
{Berger}, E., {Fong}, W., \& {Chornock}, R. 2013, \apjl, 774, L23,
  \dodoi{10.1088/2041-8205/774/2/L23}

\bibitem[{{Bromberg} {et~al.}(2011){Bromberg}, {Nakar}, \&
  {Piran}}]{Bromberg11}
{Bromberg}, O., {Nakar}, E., \& {Piran}, T. 2011, \apjl, 739, L55,
  \dodoi{10.1088/2041-8205/739/2/L55}

\bibitem[{Bromberg {et~al.}(2013)Bromberg, Nakar, Piran, \&
  Sari}]{Bromberg_2013classification}
Bromberg, O., Nakar, E., Piran, T., \& Sari, R. 2013, The Astrophysical
  Journal, 764, 179, \dodoi{10.1088/0004-637x/764/2/179}

\bibitem[{Bufano {et~al.}(2012)Bufano, Pian, Sollerman, Benetti, Pignata,
  Valenti, Covino, D’Avanzo, Malesani, Cappellaro, \& et~al.}]{Bufano_2012}
Bufano, F., Pian, E., Sollerman, J., {et~al.} 2012, The Astrophysical Journal,
  753, 67, \dodoi{10.1088/0004-637x/753/1/67}

\bibitem[{{Cano} {et~al.}(2017){Cano}, {Wang}, {Dai}, \& {Wu}}]{Cano2017}
{Cano}, Z., {Wang}, S.-Q., {Dai}, Z.-G., \& {Wu}, X.-F. 2017, Advances in
  Astronomy, 2017, 8929054, \dodoi{10.1155/2017/8929054}

\bibitem[{Cano {et~al.}(2014)Cano, de~Ugarte~Postigo, Pozanenko, Butler,
  Thöne, Guidorzi, Krühler, Gorosabel, Jakobsson, Leloudas, \&
  et~al.}]{Cano_2014}
Cano, Z., de~Ugarte~Postigo, A., Pozanenko, A., {et~al.} 2014, Astronomy \&
  Astrophysics, 568, A19, \dodoi{10.1051/0004-6361/201423920}

\bibitem[{{Della Valle} {et~al.}(2006){Della Valle}, {Chincarini}, {Panagia},
  {Tagliaferri}, {Malesani}, {Testa}, {Fugazza}, {Campana}, {Covino},
  {Mangano}, {Antonelli}, {D'Avanzo}, {Hurley}, {Mirabel}, {Pellizza},
  {Piranomonte}, \& {Stella}}]{DellaValle2006}
{Della Valle}, M., {Chincarini}, G., {Panagia}, N., {et~al.} 2006, \nat, 444,
  1050, \dodoi{10.1038/nature05374}

\bibitem[{Deng(2012)}]{deng2012mnist}
Deng, L. 2012, IEEE Signal Processing Magazine, 29, 141

\bibitem[{{Fruchter} {et~al.}(2006){Fruchter}, {Levan}, {Strolger},
  {Vreeswijk}, {Thorsett}, {Bersier}, {Burud}, {Castro Cer{\'o}n},
  {Castro-Tirado}, {Conselice}, {Dahlen}, {Ferguson}, {Fynbo}, {Garnavich},
  {Gibbons}, {Gorosabel}, {Gull}, {Hjorth}, {Holland }, {Kouveliotou}, {Levay},
  {Livio}, {Metzger}, {Nugent}, {Petro}, {Pian}, {Rhoads}, {Riess}, {Sahu},
  {Smette}, {Tanvir}, {Wijers}, \& {Woosley}}]{06environmentfruchter}
{Fruchter}, A.~S., {Levan}, A.~J., {Strolger}, L., {et~al.} 2006, \nat, 441,
  463, \dodoi{10.1038/nature04787}

\bibitem[{{Fynbo} {et~al.}(2006){Fynbo}, {Watson}, {Th{\"o}ne}, {Sollerman},
  {Bloom}, {Davis}, {Hjorth}, {Jakobsson}, {J{\o}rgensen}, {Graham},
  {Fruchter}, {Bersier}, {Kewley}, {Cassan}, {Castro Cer{\'o}n}, {Foley},
  {Gorosabel}, {Hinse}, {Horne}, {Jensen}, {Klose}, {Kocevski}, {Marquette},
  {Perley}, {Ramirez-Ruiz}, {Stritzinger}, {Vreeswijk}, {Wijers}, {Woller},
  {Xu}, \& {Zub}}]{NoSN2006Nature}
{Fynbo}, J. P.~U., {Watson}, D., {Th{\"o}ne}, C.~C., {et~al.} 2006, \nat, 444,
  1047, \dodoi{10.1038/nature05375}

\bibitem[{{Ghirlanda} {et~al.}(2009){Ghirlanda}, {Nava}, {Ghisellini},
  {Celotti}, \& {Firmani}}]{Ghirlanda2009}
{Ghirlanda}, G., {Nava}, L., {Ghisellini}, G., {Celotti}, A., \& {Firmani}, C.
  2009, \aap, 496, 585, \dodoi{10.1051/0004-6361/200811209}

\bibitem[{{Ghirlanda} {et~al.}(2018){Ghirlanda}, {Nappo}, {Ghisellini},
  {Melandri}, {Marcarini}, {Nava}, {Salafia}, {Campana}, \&
  {Salvaterra}}]{Ghirlanda2018}
{Ghirlanda}, G., {Nappo}, F., {Ghisellini}, G., {et~al.} 2018, \aap, 609, A112,
  \dodoi{10.1051/0004-6361/201731598}

\bibitem[{Goldstein {et~al.}(2017)Goldstein, Veres, Burns, Briggs, Hamburg,
  Kocevski, Wilson-Hodge, Preece, Poolakkil, Roberts, Hui, Connaughton,
  Racusin, von Kienlin, Canton, Christensen, Littenberg, Siellez, Blackburn,
  Broida, Bissaldi, Cleveland, Gibby, Giles, Kippen, McBreen, McEnery, Meegan,
  Paciesas, \& Stanbro}]{fermi170817A}
Goldstein, A., Veres, P., Burns, E., {et~al.} 2017, The Astrophysical Journal,
  848, L14, \dodoi{10.3847/2041-8213/aa8f41}

\bibitem[{{Hjorth} \& {Bloom}(2012)}]{Hjorth2012}
{Hjorth}, J., \& {Bloom}, J.~S. 2012, {The Gamma-Ray Burst - Supernova
  Connection} (Cambridge University Press), 169--190

\bibitem[{{Hjorth} {et~al.}(2003){Hjorth}, {Sollerman}, {M{\o}ller}, {Fynbo},
  {Woosley}, {Kouveliotou}, {Tanvir}, {Greiner}, {Andersen}, {Castro-Tirado},
  {Castro Cer{\'o}n}, {Fruchter}, {Gorosabel}, {Jakobsson}, {Kaper}, {Klose},
  {Masetti}, {Pedersen}, {Pedersen}, {Pian}, {Palazzi}, {Rhoads}, {Rol}, {van
  den Heuvel}, {Vreeswijk}, {Watson}, \& {Wijers}}]{hjorth2003}
{Hjorth}, J., {Sollerman}, J., {M{\o}ller}, P., {et~al.} 2003, \nat, 423, 847,
  \dodoi{10.1038/nature01750}

\bibitem[{Kann {et~al.}(2019)Kann, Schady, Olivares~E., Klose, Rossi, Perley,
  Krühler, Greiner, Nicuesa~Guelbenzu, Elliott, \& et~al.}]{Kann_2019}
Kann, D.~A., Schady, P., Olivares~E., F., {et~al.} 2019, Astronomy \&
  Astrophysics, 624, A143, \dodoi{10.1051/0004-6361/201629162}

\bibitem[{{Kann} {et~al.}(2019){Kann}, {Schady}, {Olivares E.}, {Klose},
  {Rossi}, {Perley}, {Kr{\"u}hler}, {Greiner}, {Nicuesa Guelbenzu}, {Elliott},
  {Knust}, {Filgas}, {Pian}, {Mazzali}, {Fynbo}, {Leloudas}, {Afonso},
  {Delvaux}, {Graham}, {Rau}, {Schmidl}, {Schulze}, {Tanga}, {Updike}, \&
  {Varela}}]{1SNGRB}
{Kann}, D.~A., {Schady}, P., {Olivares E.}, F., {et~al.} 2019, \aap, 624, A143,
  \dodoi{10.1051/0004-6361/201629162}

\bibitem[{Klose {et~al.}(2019)Klose, Schmidl, Kann, Nicuesa~Guelbenzu, Schulze,
  Greiner, Olivares~E., Krühler, Schady, Afonso, \& et~al.}]{Klose_2019}
Klose, S., Schmidl, S., Kann, D.~A., {et~al.} 2019, Astronomy \& Astrophysics,
  622, A138, \dodoi{10.1051/0004-6361/201832728}

\bibitem[{{Kouveliotou} {et~al.}(1993){Kouveliotou}, {Meegan}, {Fishman},
  {Bhat}, {Briggs}, {Koshut}, {Paciesas}, \& {Pendleton}}]{Kouveliotou1993}
{Kouveliotou}, C., {Meegan}, C.~A., {Fishman}, G.~J., {et~al.} 1993, \apjl,
  413, L101, \dodoi{10.1086/186969}

\bibitem[{{Lamb} {et~al.}(2019){Lamb}, {Tanvir}, {Levan}, {de Ugarte Postigo},
  {Kawaguchi}, {Corsi}, {Evans}, {Gompertz}, {Malesani}, {Page}, {Wiersema},
  {Rosswog}, {Shibata}, {Tanaka}, {van der Horst}, {Cano}, {Fynbo}, {Fruchter},
  {Greiner}, {Heintz}, {Higgins}, {Hjorth}, {Izzo}, {Jakobsson}, {Kann},
  {O'Brien}, {Perley}, {Pian}, {Pugliese}, {Starling}, {Th{\"o}ne}, {Watson},
  {Wijers}, \& {Xu}}]{Lamb2019}
{Lamb}, G.~P., {Tanvir}, N.~R., {Levan}, A.~J., {et~al.} 2019, \apj, 883, 48,
  \dodoi{10.3847/1538-4357/ab38bb}

\bibitem[{LeCun {et~al.}(1998)LeCun, Bottou, Bengio, Haffner,
  {et~al.}}]{mnistlecun1998gradient}
LeCun, Y., Bottou, L., Bengio, Y., Haffner, P., {et~al.} 1998, Proceedings of
  the IEEE, 86, 2278

\bibitem[{{Lien} {et~al.}(2016){Lien}, {Sakamoto}, {Barthelmy}, {Baumgartner},
  {Cannizzo}, {Chen}, {Collins}, {Cummings}, {Gehrels}, {Krimm}, {Markwardt},
  {Palmer}, {Stamatikos}, {Troja}, \& {Ukwatta}}]{Lien2016}
{Lien}, A., {Sakamoto}, T., {Barthelmy}, S.~D., {et~al.} 2016, \apj, 829, 7,
  \dodoi{10.3847/0004-637X/829/1/7}

\bibitem[{{Liu} {et~al.}(2019){Liu}, {Zhang}, {Howard}, {Bai}, {Lu}, {Soria},
  {Justham}, {Li}, {Zheng}, {Wang}, {Belczynski}, {Casares}, {Zhang}, {Yuan},
  {Dong}, {Lei}, {Isaacson}, {Wang}, {Bai}, {Shao}, {Gao}, {Wang}, {Niu},
  {Cui}, {Zheng}, {Mu}, {Zhang}, {Wang}, {Heger}, {Qi}, {Liao}, {Lattanzi},
  {Gu}, {Wang}, {Wu}, {Shao}, {Shen}, {Wang}, {Bregman}, {Di Stefano}, {Liu},
  {Han}, {Zhang}, {Wang}, {Ren}, {Zhang}, {Zhang}, {Wang}, {Cabrera-Lavers},
  {Corradi}, {Rebolo}, {Zhao}, {Zhao}, {Chu}, \& {Cui}}]{Liu2019}
{Liu}, J., {Zhang}, H., {Howard}, A.~W., {et~al.} 2019, \nat, 575, 618,
  \dodoi{10.1038/s41586-019-1766-2}

\bibitem[{Maaten \& Hinton(2008)}]{maaten2008visualizing}
Maaten, L. v.~d., \& Hinton, G. 2008, Journal of machine learning research, 9,
  2579

\bibitem[{{Nakar}(2007)}]{NAKARsgrb}
{Nakar}, E. 2007, \physrep, 442, 166, \dodoi{10.1016/j.physrep.2007.02.005}

\bibitem[{{Norris} \& {Bonnell}(2006)}]{Norris2006}
{Norris}, J.~P., \& {Bonnell}, J.~T. 2006, \apj, 643, 266,
  \dodoi{10.1086/502796}

\bibitem[{{Norris} {et~al.}(1986){Norris}, {Share}, {Messina}, {Dennis},
  {Desai}, {Cline}, {Matz}, \& {Chupp}}]{Norris1986}
{Norris}, J.~P., {Share}, G.~H., {Messina}, D.~C., {et~al.} 1986, \apj, 301,
  213, \dodoi{10.1086/163889}

\bibitem[{{Ofek} {et~al.}(2007){Ofek}, {Cenko}, {Gal-Yam}, {Fox}, {Nakar},
  {Rau}, {Frail}, {Kulkarni}, {Price}, {Schmidt}, {Soderberg}, {Peterson},
  {Berger}, {Sharon}, {Shemmer}, {Penprase}, {Chevalier}, {Brown}, {Burrows},
  {Gehrels}, {Harrison}, {Holland }, {Mangano}, {McCarthy}, {Moon}, {Nousek},
  {Persson}, {Piran}, \& {Sari}}]{Ofek2007}
{Ofek}, E.~O., {Cenko}, S.~B., {Gal-Yam}, A., {et~al.} 2007, \apj, 662, 1129,
  \dodoi{10.1086/518082}

\bibitem[{{Paciesas} {et~al.}(1999){Paciesas}, {Meegan}, {Pendleton}, {Briggs},
  {Kouveliotou}, {Koshut}, {Lestrade}, {McCollough}, {Brainerd}, {Hakkila},
  {Henze}, {Preece}, {Connaughton}, {Kippen}, {Mallozzi}, {Fishman},
  {Richardson}, \& {Sahi}}]{Paciesas1999}
{Paciesas}, W.~S., {Meegan}, C.~A., {Pendleton}, G.~N., {et~al.} 1999, \apjs,
  122, 465, \dodoi{10.1086/313224}

\bibitem[{Sacchi \& Ulrych(1998)}]{FTIEEEzero}
Sacchi, W., \& Ulrych. 1998, IEEE, \dodoi{10.1109/78.651165}

\bibitem[{Shen \& Wang(2006)}]{FTopticszero}
Shen, F., \& Wang, A. 2006, Appl. Opt., 45, 1102, \dodoi{10.1364/AO.45.001102}

\bibitem[{{Stanek} {et~al.}(2003){Stanek}, {Matheson}, {Garnavich}, {Martini},
  {Berlind}, {Caldwell}, {Challis}, {Brown}, {Schild}, {Krisciunas}, {Calkins},
  {Lee}, {Hathi}, {Jansen}, {Windhorst}, {Echevarria}, {Eisenstein}, {Pindor},
  {Olszewski}, {Harding}, {Holland }, \& {Bersier}}]{Stanek2003}
{Stanek}, K.~Z., {Matheson}, T., {Garnavich}, P.~M., {et~al.} 2003, \apjl, 591,
  L17, \dodoi{10.1086/376976}

\bibitem[{Steinhardt {et~al.}(2020)Steinhardt, Weaver, Maxfield, Davidzon,
  Faisst, Masters, Schemel, \& Toft}]{Steinhardt_tsne_galaxy2020}
Steinhardt, C.~L., Weaver, J.~R., Maxfield, J., {et~al.} 2020, The
  Astrophysical Journal, 891, 136, \dodoi{10.3847/1538-4357/ab76be}

\bibitem[{{Tanvir} {et~al.}(2013){Tanvir}, {Levan}, {Fruchter}, {Hjorth},
  {Hounsell}, {Wiersema}, \& {Tunnicliffe}}]{1GRBKN}
{Tanvir}, N.~R., {Levan}, A.~J., {Fruchter}, A.~S., {et~al.} 2013, \nat, 500,
  547, \dodoi{10.1038/nature12505}

\bibitem[{Tavani {et~al.}(1998)Tavani, Kniffen, Mattox, Paredes, \&
  Foster}]{Tavani1998}
Tavani, M., Kniffen, D., Mattox, J., Paredes, J., \& Foster, R. 1998, The
  Astrophysical Journal Letters, 497, L89

\bibitem[{{Troja} {et~al.}(2019){Troja}, {Castro-Tirado}, {Becerra
  Gonz{\'a}lez}, {Hu}, {Ryan}, {Cenko}, {Ricci}, {Novara},
  {S{\'a}nchez-R{\'a}mirez}, {Acosta-Pulido}, {Ackley}, {Caballero
  Garc{\'\i}a}, {Eikenberry}, {Guziy}, {Jeong}, {Lien}, {M{\'a}rquez}, {Pand
  ey}, {Park}, {Sakamoto}, {Tello}, {Sokolov}, {Sokolov}, {Tiengo}, {Valeev},
  {Zhang}, \& {Veilleux}}]{2019MNRASKN}
{Troja}, E., {Castro-Tirado}, A.~J., {Becerra Gonz{\'a}lez}, J., {et~al.} 2019,
  \mnras, 489, 2104, \dodoi{10.1093/mnras/stz2255}

\bibitem[{van~der Maaten(2015)}]{vandermaaten2015}
van~der Maaten, L. 2015, Journal of Machine Learning Research, 15, 3221

\bibitem[{Zhang {et~al.}(2012)Zhang, Shao, Yan, \& Wei}]{zhang2012revisiting}
Zhang, F.-W., Shao, L., Yan, J.-Z., \& Wei, D.-M. 2012, The Astrophysical
  Journal, 750, 88

\end{thebibliography}
\end{document}